\documentclass[10pt,aps,prc,twocolumn,floatfix]{revtex4-1} 
\usepackage{graphicx}
\usepackage{sidecap}
\usepackage{array}
\usepackage{multirow}
\usepackage{subfigure}
\usepackage{color}
\usepackage{lineno}

\begin{document}

\title{Centrality and transverse momentum dependence of elliptic flow
  of multi-strange hadrons and $\phi$ meson in Au+Au collisions at $\sqrt{s_{NN}}$ = 200 GeV }

\author{
L.~Adamczyk$^{1}$,
J.~K.~Adkins$^{20}$,
G.~Agakishiev$^{18}$,
M.~M.~Aggarwal$^{30}$,
Z.~Ahammed$^{47}$,
I.~Alekseev$^{16}$,
A.~Aparin$^{18}$,
D.~Arkhipkin$^{3}$,
E.~C.~Aschenauer$^{3}$,
G.~S.~Averichev$^{18}$
% ...
}

\affiliation{$^{1}$AGH University of Science and Technology, Cracow 30-059, Poland}
% ...
\affiliation{$^{3}$Brookhaven National Laboratory, Upton, New York 11973, USA}
% ...
\affiliation{$^{16}$Alikhanov Institute for Theoretical and Experimental Physics, Moscow 117218, Russia}
% ...
\affiliation{$^{18}$Joint Institute for Nuclear Research, Dubna, 141 980, Russia}
% ...
\affiliation{$^{20}$University of Kentucky, Lexington, Kentucky, 40506-0055, USA}
% ...
\affiliation{$^{30}$Panjab University, Chandigarh 160014, India}
% ...
\affiliation{$^{47}$Variable Energy Cyclotron Centre, Kolkata 700064, India}
% ...

\collaboration{STAR Collaboration}\noaffiliation



\begin{abstract}
We present high precision measurements of elliptic flow  near
midrapidity ($|y|<1.0$) for multi-strange hadrons and $\phi$ meson
as a function of centrality and transverse momentum in Au+Au collisions at center of mass
energy $\sqrt{s_{NN}}=$ 200 GeV. We observe that the transverse momentum dependence of $\phi$ and $\Omega$ $v_{2}$  is
similar to that of $\pi$ and $p$, respectively, which may indicate that the heavier strange  quark
flows as strongly as the lighter  up  and down 
quarks. This observation constitutes a clear piece of evidence for the development
of partonic collectivity in heavy-ion collisions at the top RHIC energy.
Number of constituent quark  scaling is found to hold within
 statistical uncertainty for both 0-30$\%$ and 30-80$\%$ collision
 centrality. There is an indication of the breakdown of previously
 observed mass ordering between $\phi$ and proton $v_{2}$ at low transverse momentum  in the 0-30$\%$ centrality range, possibly indicating late hadronic interactions affecting the proton $v_{2}$.
\end{abstract}
\pacs{25.75.-q}
\maketitle
%

At sufficiently high temperature and/or high density Quantum Chromodynamics (QCD) predicts a transition 
form hadronic matter to de-confined partonic matter~\cite{qcd}.
The main goal of the STAR (Solenoid Tracker at RHIC) experiment at the
Relativistic Heavy Ion Collider (RHIC) is to study the properties of
QCD matter at extremely high energy and parton densities, created
in the heavy-ion collision.
In high energy heavy-ion collisions, particles are produced with an
azimuthally anisotropic momentum distribution, which is a result of
hydrodynamical flow of the Quark-Gluon-Plasma (in the soft regime). One way to examine
this anisotropy is to measure elliptic flow ($v_{2}$), which plays a
crucial role in the study of
the  QCD matter formed during the collision.
The elliptic flow, defined as $v_{2}=\langle\cos 2(\varphi-\Psi)\rangle$, is the second Fourier coefficient 
of the azimuthal distribution of the emitted particle with respect to the
reaction plane (defined by the beam axis and a vector between the centers of the colliding ions). 
Here $\varphi$ is the azimuthal angle of emitted particle and $\Psi$
is the azimuthal angle of the reaction plane.
Over the past decade, experimental measurements have shown elliptic flow to be especially sensitive to the initial phase and equation of state of the system formed in heavy-ion collisions~\cite{flow1,flow2,starflow1,starflow2,starflow3}.
However, information about the early dynamics of the system  may be modified by
hadronic re-scattering in the later stage of the collision~\cite{hyrdo_cascade,hyrdo_cascade2}. The hadronic interaction cross-sections of
$\phi$, $\Xi$ and $\Omega$ are expected to be
small~\cite{multistrange1} and their freeze-out temperatures are close to the quark-hadron transition
temperature predicted by lattice QCD~\cite{white,multistrange2}. Hence, these
hadrons are expected to provide information primarily  from the
partonic stage of the
collision~\cite{bm,jinhui1,jinhui2,nbn,xzhu}.
Previous measurements of $\phi$ and $\Omega$ $v_{2}$ from STAR~\cite{v00} were statistically limited and little is known about the centrality dependence of $\Omega$ $v_{2}$.
The measurements of $\phi$ and $\Omega$ $v_{2}$ presented here as a
function of both transverse momentum ($p_{T}$) and centrality help to alleviate these limitations.
Moreover,  high precision
measurements of $\phi$-meson $v_{2}$ relative to proton
$v_{2}$ at low $p_{T}$ may provide information on  the effect of hadronic
re-scattering~\cite{hyrdo_cascade,hyrdo_cascade2} in the late stages of the collision. \\
We present the collision centrality and $p_{T}$ dependence of the
elliptic flow  of $\pi^{+}+\pi^{-}$, $K^{+}+K^{-}$, $K^{0}_{S}$, $p+\bar{p}$, $\phi$, $\Lambda+\overline{\Lambda}$, $\Xi^{-}+\overline{\Xi}^{+}$ and $\Omega^{-}+\overline{\Omega}^{+}$. 
For this study we used 730 million of  Au+Au events at $\sqrt{s_{NN}}=200$ GeV recorded by STAR in 2010 and 2011 with a minimum-bias trigger~\cite{trigger}. The collision centrality is determined by comparing the measured raw
charged hadron multiplicity from the Time Projection Chamber (TPC) within a pseudorapidity window $|\eta|<$ 0.5 with Glauber Monte-Carlo simulations~\cite{glb,centrality}.
The TPC and Time of Flight (TOF) detectors
with full  azimuthal coverage are used for particle identification in the
central rapidity region ($|\eta|<$ 1.0 for TPC and $|\eta|<$ 0.9 for
TOF). Charged particles are identified using specific ionization energy loss as a function
of momentum (in the TPC) and square of the particle mass as a function
of momentum (for the TOF). We reconstruct short-lived  $K^{0}_{S}$,
$\Lambda$, $\Xi$, $\Omega$  and $\phi$ through the following decay channels :
$K^{0}_{S}$ $\rightarrow$ $\pi^{+}$ + $\pi^{-}$, $\Lambda$
$\rightarrow$ $p$ + $\pi$, $\Xi$ $\rightarrow$ $\Lambda$ +$\pi$, 
$\Omega$ $\rightarrow$ $\Lambda$ + $\it{K}$ and $\phi$ $\rightarrow$ $\it{K}^{+}$ +
$\it{K}^{-}$. 
Topological and kinematic cuts are applied to reduce the combinatorial background
for $K^{0}_{S}$, $\Lambda$, $\Xi$ and $\Omega$. The detailed
description of the procedures can be found in
Refs.~\cite{v00,v01,v02}.\\
The $\eta$ sub-event plane method~\cite{method} is used for the elliptic flow analysis. An $\eta$ gap of $|\Delta\eta| >$ 0.1 between positive and negative
pseudorapidity sub-events is introduced to suppress non-flow
effects. The $v_{2}$ for short-lived hadrons ($K^{0}_{S}$, $\phi$, $\Lambda$, $\Xi$ and $\Omega$) is calculated as a function of
invariant mass for each $p_{T}$ and centrality bin in order to take into
account the invariant mass dependence of the signal to background ratio.
Details of this method can be found in Ref.~\cite{inv_mass_method}. The observed $v_{2}$ values are
corrected for finite event plane resolution which is determined by
comparing the two $\eta$-sub event plane angles. A resolution correction
is done by dividing the term $\cos 2(\varphi-\Psi_{2})$
by the event plane resolution for the corresponding centrality for
each event following the method described in
Refs.~\cite{res_corr,res_corr_nsm}. The change in $v_{2}$ between present
method of resolution correction and  the previous method used in earlier STAR
publication~\cite{starflow1,starflow2,starflow3} is $\leq$ 5$\%$ at  $\sqrt{s_{NN}}=200$ GeV. Here $\Psi_{2}$ is the $2^{\text{nd}}$ order
event plane which is used for $v_{2}$ measurements.\\
For all particle species, the cuts used for particle identification (PID) and
background subtraction are varied to estimate the systematic errors. 
Furthermore, different techniques (e.g. by counting entries in each bin
of the invariant mass histogram or by fitting the
shape of the invariant mass distribution using a function)  for yield extraction are used. For
$\pi^{\pm}$, $K^{\pm}$ and $p(\bar{p})$, 6 different combinations of
track cuts and 3 different sets of PID cuts which finally yield 18
combinations have been used. For other strange hadrons ($K^{0}_{S}$,
$\phi$, $\Lambda$, $\Xi$ and $\Omega$) 20 different cut combination
are used. 
The root-mean-square value of
point-by-point difference from the default value ($v_{2}$ from default
set of cuts) is used as the systematic error on each data point.
 The total systematic error depends on $p_{T}$,
centrality and particle species. We observed 3-5$\%$ systematic error
for $p_{T}$ $<$ 1.5 GeV/$c$  and  0-30$\%$ centrality for $\phi$,
$K^{0}_{S}$, $\Lambda$ whereas for $\Xi$ and $\Omega$ the systematic error varies
from  8$\%$ to 14$\%$. Total systematic errors are less than 1$\%$ for
$\pi^{\pm}$, $K^{\pm}$ and $p(\bar{p})$ for all $p_{T}$ and
centralities.\\ 
We investigated the effect of track reconstruction efficiency on the
measured $v_{2}$ of identified hadrons for wide centrality bins, such
as a 0-80$\%$ centrality bin. The centrality
dependence of track reconstruction efficiency biases the measured
$v_{2}$ toward  events with higher reconstruction efficiency, an
effect we will refer to as an ``efficiency bias".
Due to the efficiency bias, the $v_{2}$ of $\Xi$ and $\Omega$, each
having three daughters, changes by no more  than 5$\%$ in 0-80$\%$ centrality. 
For the other measured  particles, the effect is less than 3$\%$ for 0-80$\%$
centrality. The $v_{2}(p_{T})$ of all particles presented here 
have been corrected for the efficiency bias by using the inverse of efficiency
as a weight for the $v_{2}$ as a function of $p_{T}$ and centrality. \\
An additional correction is needed for $\phi$, $\Xi$ and $\Omega$ $v_{2}$.
An event bias is naturally introduced when one measures $v_{2}$ in wide centrality bins, especially for the rare particles.
As the measured $v_{2}$ is an average over all events weighted by
particle yield, the average event shape depends on the particle
type. A Glauber model~\cite{glb} study of the average initial participant eccentricity indicates the
multi-strange hadron $v_{2}$ is more biased toward central events than
that of the light and strange hadrons.
Specifically, the average eccentricity for multi-strange hadrons
in wide centrality bins  is smaller than the eccentricity determined by
the particle yield of all charged hadrons. One should take this
effect into consideration if any conclusion on the number of constituent quark scaling is drawn.
This bias can be corrected by normalizing the measured $v_{2}$ by the  ratio of
eccentricity to that weighted by the yield of the particle of interest.
We find the event bias correction factors for 0-30$\%$, 30-80$\%$,
and 0-80$\%$ centralities are
1.002, 1.053 and 1.028 for  $\phi$; 1.019, 1.054 and 1.091 for $\Xi$;  1.068,  1.067 and  1.177 for $\Omega$.
The event bias correction for light and strange
 hadrons is small ($<$ 0.03), perhaps be due to  their copious production.
 Therefore, in the later discussion of number of constituent quark (NCQ) scaling, the event bias
 correction is applied only to the $v_{2}$ of
multi-strange hadrons and $\phi$ meson. The above correction factors
remain almost unchanged if we use Color-Glass Condensate
(CGC)~\cite{cgc} based model to calculate eccentricity.\\
\begin{figure}[]
\centering
\centerline{\includegraphics[scale=0.47]{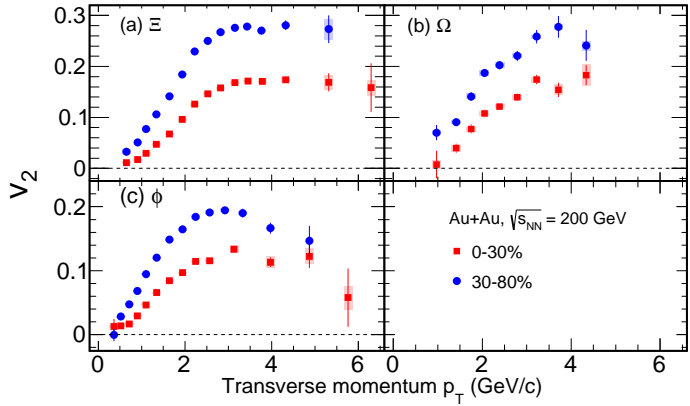}}
  \caption{(Color online) The $v_{2}$ as a function of $p_{T}$ near
    midrapidity ($|y|<1.0$) for  (a) $\Xi^{-}$ + $\overline{\Xi}^{+}$ (b)
    $\Omega^{-}$ + $\overline{\Omega}^{+}$ and (c) $\phi$ from Au + Au
    collisions at $\sqrt{s_{NN}}$ = 200 GeV for 0-30$\%$ and
    30-80$\%$ centrality. The systematic uncertainties are shown by shaded boxes and the statistical uncertainties by vertical lines. }
\label{cent}       
\end{figure}
In Figure~\ref{cent} we present the elliptic flow parameter
$v_{2}(p_{T})$ at  midrapidity ($|y|<1.0$) for  (a) $\Xi^{-}$ + $\overline{\Xi}^{+}$, (b)
$\Omega^{-}$ + $\overline{\Omega}^{+}$ and (c) $\phi$ in Au + Au collisions at $\sqrt{s_{NN}}$ = 200 GeV for 0-30$\%$ and
30-80$\%$ centrality. Event bias correction factors have been applied to the results shown in Fig~\ref{cent}.
A clear centrality dependence of $v_{2}(p_{T} )$ is observed for
$\phi$, $\Xi$ and $\Omega$, similar to that of  identified light and
strange hadrons previously  measured
by the STAR experiment~\cite{pid_star_flow}. The values of $v_{2}$ are
found to be larger in peripheral collisions (30-80$\%$ centrality) compared to those in  central collisions (0-30$\%$ centrality).
This observation is consistent with an interpretation in which the final momentum anisotropy is driven by the initial spatial anisotropy.\\
The NCQ scaling in $v_{2}$ for different identified hadrons is considered to be a good probe for studying the strongly interacting
partonic matter. 
The observed NCQ scaling of identified hadrons in experimental data ~\cite{starphiflow} indicates the importance of parton recombination in forming
hadrons in the intermediate $p_{T}$ range (2.0 GeV/c $<$ $p_{T}$ $<$ 4.0 GeV/c) ~\cite{voloshin,Fries,LXHan,lisa}. Such scaling
may indicate that collective elliptic flow develops during the
partonic phase. Previous measurements have found that $v_{2}$ of  $\pi$,
$K$, $p$, $K^{0}_{S}$, $\Lambda$, $\Xi$ and $\phi$ follow NCQ scaling
 well at top the RHIC energy ($\sqrt{s_{NN}}$ = 200 GeV)~\cite{starphiflow}. The large statistics data
 sets collected by STAR detectors allow us to measure  elliptic flow of multi-strange
hadrons, specifically that of the $\Omega$ baryon which is made of pure strange ($\it s$) or
anti-strange ($\bar{\it s}$) constituent quarks and of the $\phi$ meson, consisting of one $\it s$  and one
$\bar{\it s}$  constituent quark.\\   
\begin{figure}[]
\centering
\centerline{\includegraphics[scale=0.45]{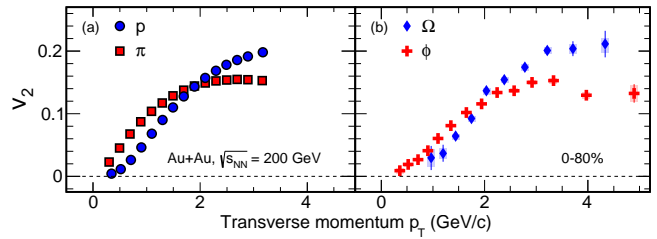}}
  \caption{(Color online) The $v_{2}$ as function of $p_{T}$ for
    $\pi$, $p$
    (panel a) and $\phi$, $\Omega$ (panel b) from minimum bias Au+Au 
    collisions at $\sqrt{s_{NN}}$ = 200 GeV for 0-80$\%$ centrality. The systematic
    uncertainties are shown by the shaded boxes while vertical lines represent the statistical uncertainties.}
\label{phi_omega}       
\end{figure}
\begin{figure*}[!ht]
\centering
\centerline{\includegraphics[scale=0.8]{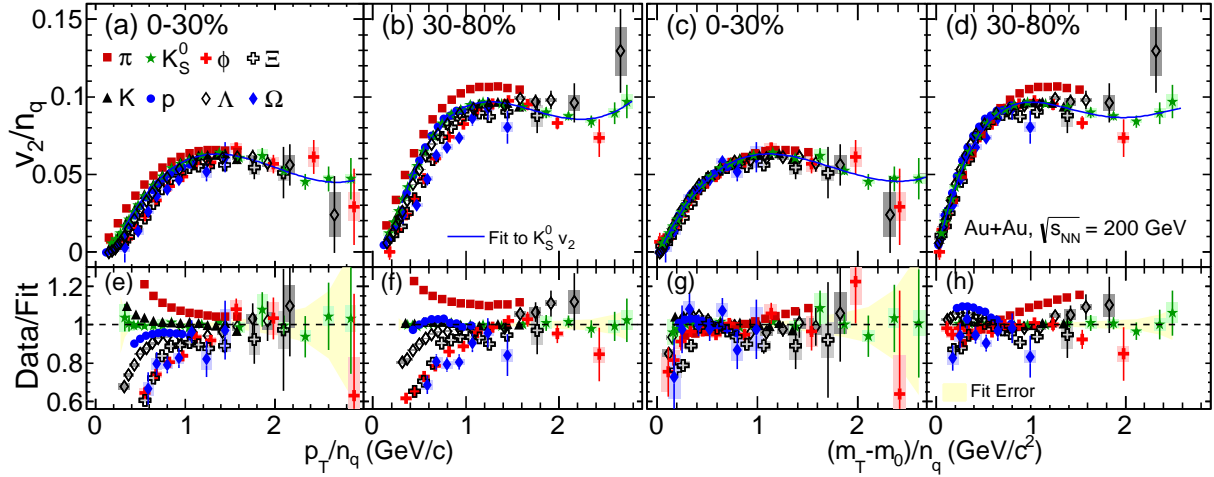}}
  \caption{(Color online) The $v_{2}$ scaled by number of constituent
    quarks ($n_{q}$) as a function of  $p_{T}/n_{q}$ and  $(m_{T}
    -m_{0})/n_{q}$ for identified hadrons from Au + Au collisions at
    $\sqrt{s_{NN}}$ = 200 GeV. Ratios with respect to a polynomial fit to the
    $K^{0}_{S}$ $v_{2}$ are shown in the corresponding lower
    panels. Vertical lines are  statistical uncertainties  and shaded boxes are
    systematic uncertainties.}
\label{phi_ncq}       
\end{figure*}
Figure~\ref{phi_omega} shows the $v_{2}$ as a function of $p_{T}$ for $\pi$, $p$, $\phi$ and $\Omega$ for 0-80$\%$ centrality in Au + Au collisions at
$\sqrt{s_{NN}}$ = 200 GeV. Here $\phi$ and $\Omega$ $v_{2}$
are corrected for the event bias mentioned earlier. Panel (a) of
Figure~\ref{phi_omega} shows a comparison between
$v_{2}$ of $\pi$ and $p$, consisting of up ($\it u$) and down ($\it d$) light quarks, and
panel (b) shows a comparison of $v_{2}$ of $\phi$ and $\Omega$
containing  heavier $\it s$ quarks. 
The $v_{2}$ of  $\phi$ and $\Omega$ are mass
ordered  at low $p_{T}$ and a baryon-meson separation  is observed at intermediate $p_{T}$.
It is clear from Figure~\ref{phi_omega} that the $v_{2}(p_{T}) $ of hadrons consisting only of strange quarks ($\phi$ and $\Omega$)  is similar to that of $\pi$ and $p$. However, unlike $\pi$ and $p$, the $\phi$ and
$\Omega$ do not participate strongly in the hadronic interactions, which suggests that the major part of collectivity is developed during the
partonic phase in Au + Au collisions at $\sqrt{s_{NN}}$ = 200 GeV.\\
We now compare our results for NCQ scaling for different collision centrality
classes to see how the partonic collectivity changes with 
system size. Figure~\ref{phi_ncq} shows the $v_{2}$ scaled by number of
constituent quarks ($n_{q}$) as a function of $p_{T}/n_{q}$ and $(m_{T}-m_{0})/n_{q}$ for
identified hadrons from Au + Au collisions at  $\sqrt{s_{NN}}$ = 200 GeV 
for 0-30$\%$ and 30-80$\%$ centrality, where $m_{T}$ and $m_{0}$ are
the transverse mass and rest mass of  hadron, respectively. Here, $\phi$, $\Xi$ and $\Omega$ $v_{2}$
are corrected for the event bias mentioned above. To quantify the deviation from
NCQ scaling, we fit the $K^{0}_{S}$ $v_{2}$ with a third-order
polynomial function.  We then take the ratio of $v_{2}$ for the other measured hadrons to the $K^{0}_{S}$ fit.  The ratios are shown in the lower panels of Figure~\ref{phi_ncq}.
 Table~\ref{ncq-table} shows the deviations of $\phi$, $\Lambda$, $\Xi$
 and $\Omega$ $v_{2}$ from the
 $K^{0}_{S}$  fit line in the range  $(m_{T}-m_{0})/n_{q}$ $>$ 0.8
 GeV/$c^{2}$.\\
\begin{table}[h]
\caption{Deviation from the
 $K^{0}_{S}$  fit line in the range  $(m_{T}-m_{0})/n_{q}$ $>$ 0.8
 GeV/$c^{2}$ for 0-30$\%$ and 30-80$\%$ centrality.}
\label{ncq-table}
\centering
\begin{center}
\scalebox{0.92}{
\begin{tabular}{|l|l|l|}
\hline
  & \multicolumn{2}{c|}{Deviation}  \\ \cline{2-3}
 Particle &   0-30$\%$  centrality        &     30-80$\%$ centrality \\ \hline
$\phi$ & 2.7$\pm$2.6($\text{stat.}$)$\pm$1.8($\text{sys.}$)$\%$  & 1.2$\pm$1.3($\text{stat.}$)$\pm$0.6($\text{sys.}$)$\%$ \\\hline
$\Lambda$ & 4.3$\pm$0.8($\text{stat.}$)$\pm$0.2($\text{sys.}$)$\%$  & 1.5$\pm$0.7($\text{stat.}$)$\pm$0.2($\text{sys.}$)$\%$ \\ \hline
 $\Xi$ & 11.3$\pm$2.3($\text{stat.}$)$\pm$1.4($\text{sys.}$)$\%$  & 8.5$\pm$2.0($\text{stat.}$)$\pm$0.5($\text{sys.}$)$\%$ \\ \hline
$\Omega$ & 10.1$\pm$8.4($\text{stat.}$)$\pm$5.3($\text{sys.}$)$\%$  & 7.0$\pm$6.0($\text{stat.}$)$\pm$1.5($\text{sys.}$)$\%$ \\ \hline
\end{tabular}}
\end{center}
\end{table}
 For both 0-30$\%$ and 30-80$\%$ centralities, the scaling holds approximately
 within 10$\%$, excluding pions. The deviation of pions could be due
 the effect of resonance decay and non-flow
 correlations~\cite{dongx}. We have seen similar order ($\sim$10$\%$) of
 deviation if we use  $p_{T}/n_{q}$ scaling as a reference. 
The maximum deviation from  NCQ scaling is $\sim$20$\%$ at $\sqrt{s_{NN}} =2.76$ TeV as observed by ALICE experiment~\cite{alicev2}. Therefore, at top RHIC
 energy, NCQ scaling holds better than LHC energy.
 The observed difference between
 the charged kaon and $K^{0}_{S}$ $v_{2}$ at low $p_{T}$ is due to
 differences in the pile-up protection conditions used in collecting
 the different data sets. The difference is taken to be an additional contribution to the systematic error on $K^{0}_{S}$ $v_{2}$.\\
\begin{figure}[!ht]
\centering
\centerline{\includegraphics[scale=0.4]{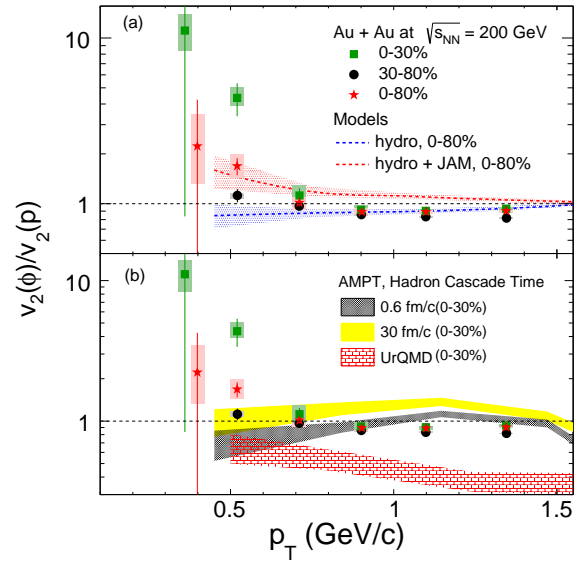}}
  \caption{(Color online) $v_{2}(\phi)/v_{2}(p)$ ratio as function of
    $p_{T}$ for  0-30$\%$, 30-80$\%$ and 0-80$\%$ centrality in Au + Au collisions at
    $\sqrt{s_{NN}}$ = 200 GeV. Shaded boxes are the  systematic uncertainties and
    vertical lines are the statistical uncertainties. The first data
    point of 0-80$\%$ centrality is shifted towards right by 400 MeV/c.
    The bands in panel
    (a) and (b) represent the hydro model~\cite{hyrdo_cascade2} and transport model calculations for
    $v_{2}(\phi)/v_{2}(p)$, respectively.  
   }
\label{phi_pr_ratio}       
\end{figure}
Hydrodynamical model calculations predict that $v_{2}$ as a function
of $p_{T}$ follows
mass ordering, where the $v_{2}$ of heavier hadrons is lower than that
of lighter hadrons and vice-versa ~\cite{flow2,hyrdo2,hyrdo3}. Mass
ordering is indeed observed in the identified hadron $v_{2}$ measured in the
low $p_{T}$ region ($p_{T}$ $\le$ 1.5 GeV/$c$)~\cite{pid_star_flow}.
Recent phenomenological calculations, based on ideal hydrodynamics together with a
hadron cascade (JAM), show that the mass ordering of $v_{2}$ could be broken
between $\phi$ mesons and protons at low $p_{ T}$ ($p_{T}$ $<$
1.5 GeV/$c$)~\cite{hyrdo_cascade,hyrdo_cascade2}.
The broken mass ordering is thought to be due to late stage hadronic
re-scattering effects on the proton $v_{2}$, since the model
calculations assume a low hadronic cross-section for the $\phi$ but a
large hadronic cross-section for the proton.\\
The ratios of $\phi$ $v_{2}$ and  proton $v_{2}$ are shown in
Figure~\ref{phi_pr_ratio}. The ratios are larger than unity at $p_{T}$
$\sim$ 0.5 GeV/$c$  for 0-30$\%$ centrality  showing an indication of breakdown of the expected mass ordering in that
momentum range. This could be due to a large  effect of hadronic
re-scattering on the proton $v_{2}$, indicated by the shaded red band in  panel (a) of
Figure~\ref{phi_pr_ratio}. We have also considered  the effect of the
momentum resolution and energy loss of the TPC as well as decay
(feed-down) effects on the proton $v_{2}$. Our study, based
on the UrQMD framework, indicates that the momentum resolution and decay
effects on the ratio of $v_{2}$($\phi$) to $v_{2}$($p$) in the
measured momentum region are negligible. The
break down of mass ordering of $v_{2}$ is more pronounced in 0-30$\%$
than in 30-80$\%$ centrality. For example, the ratio $v_{2}(\phi)$/$v_{2}(p)$ is $ 4.35 \pm 0.98 \pm^{ 0.66}_{0.45}$ at $p_{T}$ = 0.52 GeV/$c$ in 0-30$\%$,
while it is  $ 1.12 \pm 0.10 \pm^{0.047}_{0.053}$   in 30-80$\%$.
In the central
events,  both hadronic and partonic interactions are larger than in
peripheral events.
Therefore, the combined effects of large partonic collectivity on the $\phi$ $v_{2}$ and significant late stage hadronic interactions on the proton $v_{2}$ produce a greater breakdown of mass ordering in the 0-30$\%$ centrality data than in the 30-80$\%$~\cite{nbn}.
This observation indirectly supports the idea of a small hadronic
interaction cross-section for the $\phi$ meson.  We have also studied the
ratio of $\phi$ $v_{2}$ to proton $v_{2}$ using the transport
models AMPT~\cite{ampt} and UrQMD~\cite{urqmd}. The
$v_{2}(\phi)/v_{2}(p)$ ratio for 0-30$\%$ centrality from AMPT
and UrQMD model are shown in Figure~\ref{phi_pr_ratio} (panel b).
The black shaded
band is from AMPT with a hadronic cascade time of 0.6 fm/c while the yellow
band is for a hadronic cascade time of 30 fm/c. It is clear from 
Figure~\ref{phi_pr_ratio} (panel b) that with increasing hadronic cascade time
(and therefore more hadronic re-scattering), the $v_{2}(\phi)/v_{2}(p)$ ratio
increases. This is attributed to a decrease in the proton $v_{2}$ due to an  increase in hadronic re-scattering
while the $\phi$-meson $v_{2}$ remains unaffected~\cite{nbn}. The ratios from UrQMD
are shown as a red shaded band which is much smaller than unity.
The UrQMD model lacks partonic collectivity and therefore does not fully develop the $\phi$-meson $v_{2}$.\\
In summary, we have reported  high-statistics  elliptic flow measurements for multi-strange hadrons
($\Xi$ and $\Omega$) and  $\phi$ meson with other light and strange hadrons
($\pi$, $K$, $K^{0}_{S}$, $p$ and $\Lambda$) in Au + Au
collisions at $\sqrt{s_{NN}}$ = 200 GeV for different centralities. The $p_{T}$ dependence of 
$\phi$ and $\Omega$ $v_{2}$ is observed to be similar to that of $\pi$ and $p$,
indicating that a large amount of collectivity is developed in the initial
partonic phase for light and strange hadrons. NCQ scaling holds within
the statistical uncertainty for both 0-30$\%$ and  30-80$\%$
centralities, suggesting  collective motion of quarks  prior to
hadronization. 
The comparison between the $\phi$ and $p$ $v_{2}$ shows that at low
$p_{T}$, there is a possible violation of hydrodynamics-inspired mass ordering between $\phi$ and
$p$. Model calculations suggest that the
$p_{T}$ dependence of $v_{2}(\phi)/v_{2}(p)$ can be qualitatively explained by
the effect of late-stage hadronic re-scattering on the
proton $v_{2}$~\cite{hyrdo_cascade,hyrdo_cascade2}.\\

We thank the RHIC Operations Group and RCF at BNL, the NERSC Center at LBNL, the KISTI Center in
 Korea, and the Open Science Grid consortium for providing resources and support. This work was 
supported in part by the Office of Nuclear Physics within the U.S. DOE Office of Science,
 the U.S. NSF, the Ministry of Education and Science of the Russian Federation, NNSFC, CAS,
 MoST and MoE of China, the Korean Research Foundation, GA and MSMT of the Czech Republic,
 FIAS of Germany, DAE, DST, and UGC of India, the National Science Centre of Poland, National
 Research Foundation, the Ministry of Science, Education and Sports of the Republic of Croatia,
 and RosAtom of Russia.



\end{document}